\documentclass[a4paper,11pt]{article}
\usepackage{jheppub} 
\usepackage[utf8]{inputenc}
\usepackage{physics}
\usepackage{slashed}
\usepackage{caption}
\usepackage{xcolor}
\usepackage{comment}
\usepackage{multirow}
\usepackage{graphics}
\usepackage{float}
\usepackage{cases}
\usepackage{cancel}
\usepackage{soul}
\usepackage{array}
\usepackage{mathtools}  
\usepackage{amsfonts}
\usepackage{hyperref}
\usepackage{amsmath}
\usepackage{amssymb}
\usepackage{tcolorbox}
\usepackage{tikz}  
\usepackage[compat=1.0.0]{tikz-feynman}

\title{Cosmological Cutting Rules from Flat-Space Unitarity via Dressing}

\author{Arhum Ansari, Sachin Jain and Deep Mazumdar}

\affiliation{Indian Institute of Science Education and Research,\\ Dr Homi Bhabha Road, Pashan, Pune, India}

\emailAdd{ansari.arhum@students.iiserpune.ac.in,\\sachin.jain@iiserpune.ac.in\\deepkamal.mazumdar@students.iiserpune.ac.in}

\abstract{Using cosmological dressing rules, we uplift flat-space unitarity cuts to discontinuity relations for dS/EAdS observables. In this representation, Cutkosky delta functions map directly to 
“Disc”  operations in the exchanged energy variable. This provides a transparent diagram-by-diagram origin of cosmological cutting rules. We illustrate this with explicit examples at tree level and one loop for conformally coupled scalars.}

\begin{document}
\maketitle

\section{Introduction}\label{sec:intro}
Understanding how fundamental principles such as unitarity and causality manifest across different spacetime backgrounds is a central theme in quantum field theory. In flat-space, unitarity plays a fundamental role in constraining allowed class  of QFTs \cite{Peskin:1995ev,Schwartz:2014sze}. Unitarity relates the discontinuities of scattering amplitudes to sums over on-shell intermediate states. These rules provide a powerful organizational principle for loop amplitudes and underpin modern developments in amplitude-based methods. Extending this intuition to curved spacetimes remains a subtle problem, particularly in de Sitter (dS) space, where conventional notions of asymptotic states and scattering amplitudes are absent, see \cite{Marolf:2012kh}.

In the dS context, physical information is instead encoded in boundary correlation functions, whose analytic structure carries imprints of bulk dynamics. A number of works have shown that unitarity in cosmology can be packaged into discontinuity relations: at the level of the wavefunction this appears as a cosmological analogue of the optical theorem  \cite{Melville:2021lst,Goodhew:2020hob,AguiSalcedo:2023nds,Goodhew:2021oqg}, while at the correlator level one finds cutting rules that relate discontinuities to products of lower-point building blocks \cite{Das:2025qsh,Colipi-Marchant:2025oin}\footnote{{See\,\cite{Goodhew:2024eup,Thavanesan:2025kyc} for a discussion on the cosmological CPT theorem.}}. At the same time, the correlator story contains an important aspect: the objects that naturally multiply on the right-hand side are not always the correlators themselves, but particular combinations  that do not appear as standalone observables.

Independently, cosmological dressing rules \cite{Chowdhury:2024snc} provide a direct map from flat-space Feynman integrands to dS/EAdS observables by replacing propagators with simple, dressed kernels. This map  is precisely the sort of framework in which one might hope to import flat-space unitarity almost mechanically.

Through this note, we add to this story and make connection between these two threads explicit: we show how the familiar flat-space statement “unitarity = Cutkosky cuts” becomes, after dressing and analytic continuation, the discontinuity operations that appear in dS/EAdS  correlators. Put differently, we explain how the on-shell delta functions arising from Cutkosky rules uplift into the relevant cosmological “Disc” operations in the exchanged energy variable, and how the auxiliary combinations required at correlator level arise naturally in the dressed representation.


Our goal is to provide a derivation of the known cosmological cutting rules that is more closely aligned with the reasoning and techniques used in flat-space amplitude analyses. This produces a clean operational recipe: start from the flat-space optical theorem, express the cut in the standard on-shell language, and then dress the result to obtain the corresponding dS/EAdS discontinuity relation. We illustrate this dictionary on representative exchange and loop diagrams and comment on extensions (including spinning exchange).

The paper is organised in the following manner. In section \ref{Background}, we review the notions of unitarity in flat as well dS space. We describe the statement of optical theorem in flat and describe the discontinuity relations for cosmological correlators. We also review the dressing rules for conformally coupled scalars in dS (EAdS) space as given in \cite{Chowdhury:2025ohm}. In section \ref{Alt-deri}, we use split representation and flat space analouge of cutting, to derive the cosmological cutting rules. Finally in section \ref{sec:lifting}, we describe how flat-space optical theorem, with appropriate analytic continuation, can be dressed to give discontinuity relations in dS/EAdS. We demonstrate this using tree and loop level processes involving conformally coupled scalars. We finally conclude with some future directions in section \ref{sec:conc}.

\section{Background}\label{Background}
In this section, we review some of the background needed for the later part of the paper. This includes the notions of unitarity in Minkowski space and dS space, and the dressing rules. Let us discuss these one by one starting from unitarity in fat space. 

\subsection{Unitarity in flat space}\label{ssec:flat-space unitarity}
Unitarity of flat space scattering amplitude implies 
\begin{align}
    SS^\dagger=\mathbf{I}\;.
\end{align}
Writing $S=\mathbf{I}+i\mathcal{T}$, this translates to 
\begin{align}\label{eq:TTdagger_rel}
    -i(\mathcal{T}-\mathcal{T}^\dagger)=\mathcal{T}\mathcal{T}^\dagger\;.
\end{align}
We can also write \eqref{eq:TTdagger_rel} in terms of matrix elements as follows \cite{Peskin:1995ev},
\begin{align}\label{eq:matrix form TT generic}
    -i[\mathcal{A}(a\rightarrow b)-\mathcal{A}^*(a\rightarrow b)]=\sum_fd\Pi_f \mathcal{A}^*(b\rightarrow f)\mathcal{A}(a\rightarrow f),
\end{align}
where $a$ and $b$ are asymptotic states and $f$ is all possible intermediate states and $\Pi_f$ is the phase space measure. The equation \eqref{eq:matrix form TT generic} is a statement of \textbf{optical theorem}. For the case, when $a$ and $b$ are two particle states, we get
\begin{align}\label{eq:matrix for TT 2 part}
     -i[\mathcal{A}(k_1k_2\rightarrow p_1p_2)-\mathcal{A}^*(p_1p_2\rightarrow k_1k_2)]=\sum_n \Big(\prod_{i=1}^n\int\frac{d^3 q_i}{2 |\vec{q}_i|} \Big)&\mathcal{A}^*(p_1p_2\rightarrow q_i)\mathcal{A}(k_1k_2\rightarrow q_i)\nonumber\\
     &\times (2\pi)^4\delta^4(k_1+k_2-\{q_i\})\;.
\end{align}

\subsection{Unitarity in de Sitter space}\label{ssec:dS space unitarity}
In dS space, unitarity relates higher point wavefunction coefficients to lower point coefficients known as the Cosmological Optical Theorem (COT) \cite{Melville:2021lst,Goodhew:2020hob}. The COT gives rise to a set of cutting rules for wavefunction coefficients. These cutting rules states that, the wavefunction coefficient ($\psi$) factorizes into lower point wavefunction coefficients under the operation of certain discontinuities. Consider the case of four point wavefunction coefficient. A action of discontinuity operation for s-channel results in the following factorization
\begin{align}\label{discwfc}
i\textrm{Disc}_{p_s}\Big(i\psi_4(k_1,k_2,k_3,k_4)\Big)=\int_{qq'}\Big(i\textrm{Disc}_q{i\psi_3(k_1,k_2,q)}\Big)P_{qq'}\Big(i\textrm{Disc}_{q'}{i\psi_3(q',k_3,k_4)}\Big),
\end{align}
where the ``Disc" operation is defined as
\begin{align}
&\textrm{Disc}_{k_i}f(k_1,\cdots,k_i,\cdots,k_n,\{p\},\{\mathbf{k}\})\notag\\
&\equiv f(k_1,\cdots,k_i,\cdots,k_n,\{p\},\{\mathbf{k}\})-f^*(-k_1,\cdots,k_i,\cdots,-k_n,\{p\},\{-\mathbf{k}\}),
\end{align}
and $P_{qq'}$ is the power spectrum.

Applying discontinuity  rules to in-in correlator \cite{Das:2025qsh} using Schwinger-Keldysh formalism for polynomial interaction one obtains,
\begin{align}\label{disccorr}
\mathrm{Disc}_p \mathcal{B}^{(2)}(\{k_L,k_R\};p)
=
\frac{1}{2 P_p(\eta_0)}
\Big(&
\mathrm{Disc}_p \mathcal{B}^{(1)}(\{k_L\},p)
\times
\mathrm{Disc}_p {\mathcal{B}}^{(1)}(\{k_R\},p)\notag\\
&-
\widetilde{
\mathrm{Disc}}_p \tilde{\mathcal{B}}^{(1)}(\{k_L\},p)
\times
\widetilde{
\mathrm{Disc}}_p \tilde{\mathcal{B}}^{(1)}(\{k_R\},p)
\Big),
\end{align}
$\mathcal{B}^{(2)}$ is the two-site scalar correlator
and $\mathcal{B}^{(1)}$ and $\tilde{\mathcal{B}^{(1)}}$ are defined as
\begin{align}
&\mathcal{B}^{(1)}(k_1,\cdots,k_n)=\frac{2\mathbb{R}e\;\psi_n(k_1,\cdots,k_n)}{\prod_{i=1}^n2\mathbb{R}e\;\psi_2(k_i)},\notag\\
&\tilde{\mathcal{B}}^{(1)}(k_1,\cdots,k_n)=\frac{2i\mathbb{I}m\;\psi_n(k_1,\cdots,k_n)}{\prod_{i=1}^n2\mathbb{R}e\;\psi_2(k_i)}.
\end{align} 
Note that there are two types of discontinuity defined in \eqref{disccorr}
\begin{align}\label{eq:nil not}
\textrm{Disc}_p\mathcal{B}^{(1)}(\{k_L\},p)&=\mathcal{B}^{(1)}(\{k_L\},p)-\mathcal{B}^{(1)}(\{k_L\},-p) ,\notag\\
\widetilde{\textrm{Disc}}_p\mathcal{B}^{(1)}(\{k_L\},p)&=\mathcal{B}^{(1)}(\{k_L\},p)+\mathcal{B}^{(1)}(\{k_L\},-p).
\end{align}
The notation used in equation \eqref{eq:nil not} is given in more detail in equation (4.21) and figure 2 of  \cite{Das:2025qsh}.

With this prelude to cutting rules in cosmology, for both the wavefunction coefficients \eqref{discwfc} and correlators \eqref{disccorr}, we will now try delve into the origins of such rules for the latter and see if we can learn something from flat space cutting rules \cite{Cutkosky:1960sp}. To that end, let us try to understand how the cosmological correlators can be written by `dressing' the flat space amplitudes.
\subsection{Dressing rules}\label{sec:dressing}

Recently, there has been significant progress in evaluating cosmological correlators using EAdS formalism \cite{Sleight:2021plv,Chowdhury:2023arc,Chowdhury:2024snc,Chowdhury:2025ohm,MdAbhishek:2025dhx,Chowdhury:2025nnk,Ghosh:2014kba}. It was illustrated in \cite{Chowdhury:2025ohm}, that there exists a remarkable connection between the EAdS Witten diagram, and a flat-space Feynman diagram, when the latter is dressed with a theory dependent factor \cite{Chowdhury:2025ohm}. We shall briefly review this point in this section.

\subsection*{Tree level scalar correlators}
Let us start with the cubic theory of conformally coupled scalar in the EAdS shadow action.

\subsubsection*{$\mathbf{\phi}^3$ theory}
\begin{figure}[h]
    \centering
    \begin{minipage}{0.45\textwidth}
        \centering
        \begin{tikzpicture}
            \begin{feynman}
                \def\radius{2.5cm}
                \def\vsep{1.0cm}
                \draw[thick] (0,0) circle (\radius);
                \vertex (v1) at (-\vsep, 0);
                \vertex (v2) at (\vsep, 0);
                \vertex (k1) at (135:\radius);
                \vertex (k2) at (225:\radius);
                \vertex (k3) at (315:\radius);
                \vertex (k4) at (45:\radius);

                \diagram* {
                    (k1) -- [scalar] (v1),
                    (k2) -- [scalar] (v1),
                    (k4) -- [scalar] (v2),
                    (k3) -- [scalar] (v2),
                    (v1) -- [plain, edge label={$p, \vec{s}$}] (v2),
                };

                \node[anchor=south east] at (k1) {$\vec{k}_1$};
                \node[anchor=north east] at (k2) {$\vec{k}_2$};
                \node[anchor=north west] at (k3) {$\vec{k}_3$};
                \node[anchor=south west] at (k4) {$\vec{k}_4$};
            \end{feynman}
        \end{tikzpicture}
        \par \vspace{0.5em} 
        (a) (+) exchange
    \end{minipage}
    \hfill 
    \begin{minipage}{0.45\textwidth}
        \centering
        \begin{tikzpicture}
            \begin{feynman}
                \def\radius{2.5cm}
                \def\vsep{1.0cm}
                \draw[thick] (0,0) circle (\radius);
                \vertex (v1) at (-\vsep, 0);
                \vertex (v2) at (\vsep, 0);
                \vertex (k1) at (135:\radius);
                \vertex (k2) at (225:\radius);
                \vertex (k3) at (315:\radius);
                \vertex (k4) at (45:\radius);

                \diagram* {
                    (k1) -- [scalar] (v1),
                    (k2) -- [scalar] (v1),
                    (k4) -- [scalar] (v2),
                    (k3) -- [scalar] (v2),
                    (v1) -- [scalar, edge label={$p, \vec{s}$}] (v2),
                };

                \node[anchor=south east] at (k1) {$\vec{k}_1$};
                \node[anchor=north east] at (k2) {$\vec{k}_2$};
                \node[anchor=north west] at (k3) {$\vec{k}_3$};
                \node[anchor=south west] at (k4) {$\vec{k}_4$};
            \end{feynman}
        \end{tikzpicture}
        \par \vspace{0.5em}
        (b) (-) exchange
    \end{minipage}

    \caption{Witten diagrams representing 4-point correlation functions. (a) Exchange via solid (+) propagator. (b) Exchange via a dashed ($-$) propagator.}
    \label{fig:witten_diagrams}

\end{figure}
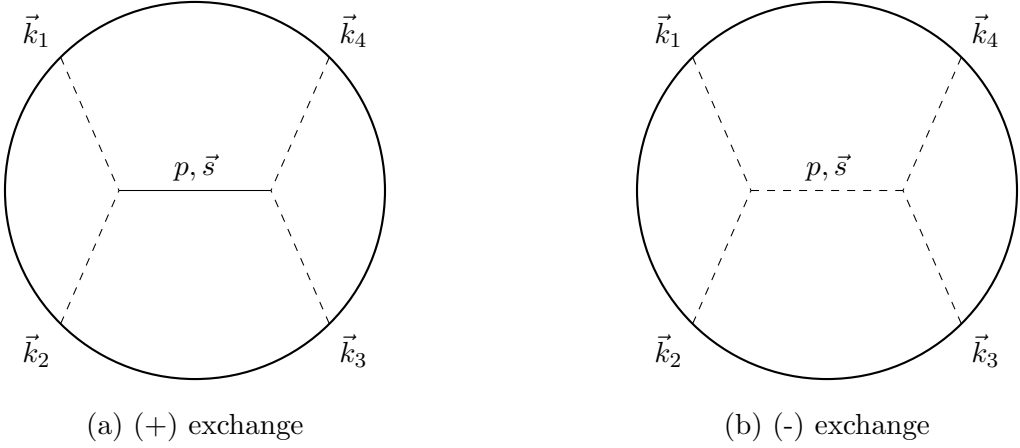

Consider the s-channel exchange diagram shown in Figure \ref{fig:witten_diagrams}, for which there can be two possible exchanges: (+) or ($-$), from $\phi_-^2\phi_+$ and $\phi_-^3$ vertices respectively \cite{Chowdhury:2025ohm}.

The Witten diagram calculation corresponding to the (+) exchange is as follows
\begin{align}\label{phi3tree+}
\left\langle \phi_1(k_1) \phi_2(k_2) \phi_3(k_3) \phi_4(k_4) \right\rangle_{s}^+=&g^2\int_0^\infty \frac{dz_1}{z_1^4}\frac{dz_2}{z_2^4} (z_1^2e^{-k_{12}z_1})(z_2^2e^{-k_{34}z_2})\int_{-\infty}^\infty\frac{pdp}{p^2+s^2}(z_1z_2)^\frac{3}{2}J_\frac{1}{2}(pz_1)J_\frac{1}{2}(pz_2)\notag\\
=&g^2\int_{-\infty}^\infty \frac{dp}{p^2+s^2}\Bigg(\int_0^\infty dz_1 \frac{e^{-k_{12}z_1}\textrm{sin}(pz_1)}{z_1} \Bigg)\Bigg(\int_0^\infty dz_2 \frac{e^{-k_{34}z_2}\textrm{sin}(pz_2)}{z_2} \Bigg),
\end{align}
where $k_{ij}=|\vec{k}_i|+|\vec{k}_j| \;\textrm{and}\; s=|\vec{k}_1+\vec{k}_2|$. The terms inside brackets are defined as the dressing factors \cite{Chowdhury:2025ohm}. Notice here that the $\textrm{sin}(pz_i)$ factors of the EAdS propagator \cite{Raju:2011mp} has been absorbed inside the dressing factor, while the remaining piece is nothing but the flat space propagator in spectral representation.

One can repeat this procedure for ($-$) exchange to obtain
\begin{align}\label{phi3tree-}
&\left\langle \phi_1(k_1) \phi_2(k_2) \phi_3(k_3) \phi_4(k_4) \right\rangle_{s}^-=-g^2\int_{-\infty}^\infty \frac{dp}{p^2+s^2}\Bigg(\int_0^\infty dz_1 \frac{e^{-k_{12}z_1}\textrm{cos}(pz_1)}{z_1} \Bigg)\Bigg(\int_0^\infty dz_2 \frac{e^{-k_{34}z_2}\textrm{cos}(pz_2)}{z_2} \Bigg),
\end{align}
where the terms in brackets are the corresponding dressing factor. We now move on to the cases involving loop diagrams.

\subsection*{Loops for scalar correlators}
In this section, we will discuss the loop diagrams for conformally coupled scalars in $\phi^4$ and $\phi^3$ theories, starting with the conformally coupled $\phi^4$ theory.

\subsubsection*{$\mathbf{\phi}^4$ theory}
Consider the four-point one-loop diagram for $\phi^4$ theory in EAdS shadow formalism. 

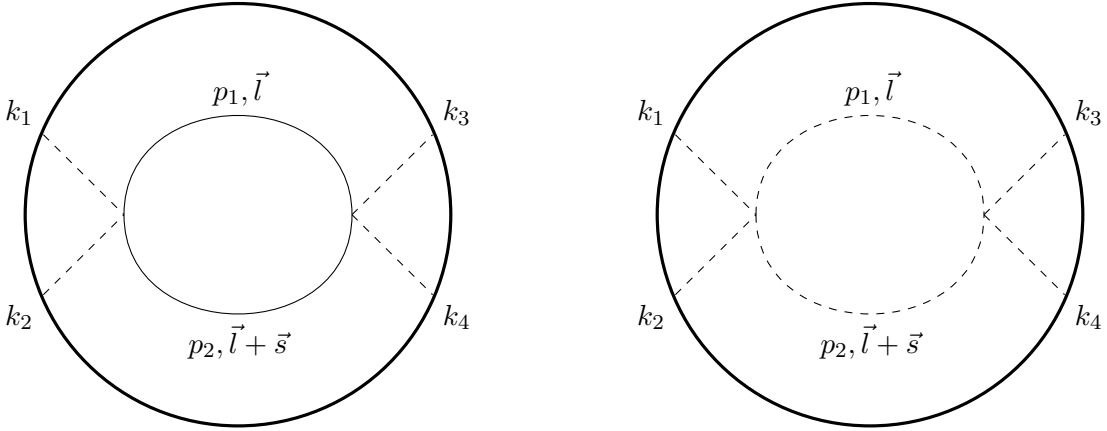
\begin{figure}[ht]
    \centering
    
    \begin{minipage}{0.45\textwidth}
        \centering
        \begin{tikzpicture}
            \begin{feynman}
                \vertex (v1);
                \vertex [right=3cm of v1] (v2);
                \vertex [above left=1.5cm of v1] (k1);
                \vertex [below left=1.5cm of v1] (k2);
                \vertex [above right=1.5cm of v2] (k3);
                \vertex [below right=1.5cm of v2] (k4);

                \diagram* {
                    (k1) -- [scalar] (v1),
                    (k2) -- [scalar] (v1),
                    (v2) -- [scalar] (k3),
                    (v2) -- [scalar] (k4),

                    (v1) -- [plain, half left, edge label={$p_1,\vec{l}$}] (v2),
                    (v1) -- [plain, half right, edge label'={$p_2,\vec{l}+\vec{s}$}] (v2),
                };
\draw[very thick]
  ($(k1)!0.5!(k4)$) circle [radius=2.8cm];
                \node[anchor=south east] at (k1) {$k_1$};
                \node[anchor=north east] at (k2) {$k_2$};
                \node[anchor=south west] at (k3) {$k_3$};
                \node[anchor=north west] at (k4) {$k_4$};
            \end{feynman}
        \end{tikzpicture}
        \par \vspace{1em}
        (a) Loop diagram  from (++) exchange
    \end{minipage}
    \hfill 
    \begin{minipage}{0.45\textwidth}
        \centering
        \begin{tikzpicture}
            \begin{feynman}
                \vertex (v1);
                \vertex [right=3cm of v1] (v2);
                \vertex [above left=1.5cm of v1] (k1);
                \vertex [below left=1.5cm of v1] (k2);
                \vertex [above right=1.5cm of v2] (k3);
                \vertex [below right=1.5cm of v2] (k4);

                \diagram* {
                    (k1) -- [scalar] (v1),
                    (k2) -- [scalar] (v1),
                    (v2) -- [scalar] (k3),
                    (v2) -- [scalar] (k4),

                    (v1) -- [scalar, half left, edge label={$p_1,\vec{l}$}] (v2),
                    (v1) -- [scalar, half right, edge label'={$p_2,\vec{l}+\vec{s}$}] (v2),
                };
\draw[very thick]
  ($(k1)!0.5!(k4)$) circle [radius=2.8cm];
                \node[anchor=south east] at (k1) {$k_1$};
                \node[anchor=north east] at (k2) {$k_2$};
                \node[anchor=south west] at (k3) {$k_3$};
                \node[anchor=north west] at (k4) {$k_4$};
            \end{feynman}
        \end{tikzpicture}
        \par \vspace{1em}
        (b) Loop diagram  from ($--$) exchange
    \end{minipage}

    \caption{One-loop Witten diagram for four point function in $\phi^4$ theory. (a) (++) exchange (b) ($--$) exchange.}
    \label{fig:1 loop four point}
    
\end{figure}

The internal propagators can be either (++) or ($--$), corresponding to $\phi_-^4$ or $\phi_-^2\phi_+^2$ vertices respectively as shown in Figure \ref{fig:1 loop four point} \cite{Chowdhury:2025ohm}. Let us start with the (++) exchange case, where the Witten diagram computation is as follows
\begin{align}\label{phi4loop+}
\left\langle \phi_1(k_1) \phi_2(k_2) \phi_3(k_3) \phi_4(k_4) \right\rangle_{s}^{++}
=&\lambda^2\int d^3l\int_{-\infty}^\infty \frac{dp_1}{(p_1^2+l^2)}\int_{-\infty}^\infty \frac{dp_2}{(p_2^2+(\vec{l}+\vec{s})^2)}\notag\\&\int_0^\infty dz_1 e^{-k_{12}z_1} \textrm{sin}(p_1z_1) \textrm{sin}(p_2z_1)\int_0^\infty dz_2e^{-k_{34}z_2}\textrm{sin}(p_1z_2)\textrm{sin}(p_2z_2),
\end{align}
while the ($--$) exchange has the following representation
\begin{align}\label{phi4loop-}
\left\langle \phi_1(k_1) \phi_2(k_2) \phi_3(k_3) \phi_4(k_4) \right\rangle_{s}^{--}
=&\lambda^2\int d^3l\int_{-\infty}^\infty \frac{dp_1}{(p_1^2+l^2)}\int_{-\infty}^\infty \frac{dp_2}{(p_2^2+(\vec{l}+\vec{s})^2)}\notag\\&\int_0^\infty dz_1 e^{-k_{12}z_1} \textrm{cos}(p_1z_1) \textrm{cos}(p_2z_1)\int_0^\infty dz_2e^{-k_{34}z_2}\textrm{cos}(p_1z_2)\textrm{cos}(p_2z_2).
\end{align}
Notice that the first line in both ($\pm$) exchanges resemble the flat-space loop integral with the energy conservation relaxed, while the second line are the dressing factors.

\subsubsection*{$\mathbf{\phi}^3$ theory}
A similar feature is observed for the two-point function in conformally coupled $\phi^3$ theory, where the allowed three point couplings include $\phi_-^3,\;\phi_-^2\phi_+$ and $\phi_-\phi_+^2$ \cite{Chowdhury:2025ohm}. We enlist the expressions for all the allowed exchanges\footnote{We have suppressed the factors of $\epsilon$ in the vertices, when compared with \cite{Chowdhury:2024snc}. These factors cancel out when the $z_i$ integrals are done appropriately with dimensional regularization.}
\begin{align}\label{phi3loop}
\langle\phi(k_1)\phi(k_2)\rangle^{++}=&+\mathfrak{I}(s)(\pi\lambda)^2\int_0^\infty\frac{dz_1}{z_1}e^{-k_{1}z_1} \textrm{sin}(p_1z_1) \textrm{sin}(p_2z_1)\int_0^\infty dz_2e^{-k_{2}z_2}\textrm{sin}(p_1z_2)\textrm{sin}(p_2z_2),\notag\\
\langle\phi(k_1)\phi(k_2)\rangle^{+-}=&-\mathfrak{I}(s)(-2\lambda)^2\int_0^\infty\frac{dz_1}{z_1}e^{-k_{1}z_1} \textrm{sin}(p_1z_1) \textrm{cos}(p_2z_1)\int_0^\infty dz_2e^{-k_{2}z_2}\textrm{sin}(p_1z_2)\textrm{cos}(p_2z_2),\notag\\
\langle\phi(k_1)\phi(k_2)\rangle^{-+}=&-\mathfrak{I}(s)(-2\lambda)^2\int_0^\infty\frac{dz_1}{z_1}e^{-k_{1}z_1} \textrm{cos}(p_1z_1) \textrm{sin}(p_2z_1)\int_0^\infty dz_2e^{-k_{2}z_2}\textrm{cos}(p_1z_2)\textrm{sin}(p_2z_2),\notag\\
\langle\phi(k_1)\phi(k_2)\rangle^{--}=&+\mathfrak{I}(s)(\pi\lambda)^2\int_0^\infty\frac{dz_1}{z_1}e^{-k_{1}z_1} \textrm{cos}(p_1z_1) \textrm{cos}(p_2z_1)\int_0^\infty dz_2e^{-k_{2}z_2}\textrm{cos}(p_1z_2)\textrm{cos}(p_2z_2).
\end{align}
where we define the flat-space loop integral as
\begin{align}
\mathfrak{I}(s)=\int d^3l\int_{-\infty}^\infty \frac{dp_1}{(p_1^2+l^2)}\int_{-\infty}^\infty \frac{dp_2}{(p_2^2+(\vec{l}+\vec{s})^2)}.
\end{align}
Thus far, we have observed how the cosmological correlators can be obtained from the flat-space amplitudes via appropriate dressing factor. We will now discuss briefly about an alternate derivation of discontinuity relations.

\section{A different derivation of the cosmological cutting rules}\label{Alt-deri}
In this section, we discuss the notion of discontinuity that will be useful in the later parts of the paper. This notion is very useful and is often employed in studing unitarity in flat space. We will show that it is equivalent to the notion of discontinuity used in the literature for dS space \cite{Melville:2021lst} and also discussed in section \ref{ssec:dS space unitarity}.

Shadow formalism is often employed to compute both the wavefunction coefficients and the in-in correlator. In shadow formalism, the $\pm$ bulk to bulk propagator for a generic scalar is given as follows \cite{DiPietro:2021sjt,Raju:2011mp},
\begin{align}\label{eq:BtB_disp_form}
    G_{\pm\nu}(s,z_1,z_2)=\int_0^\infty dp^2\frac{1}{p^2+s^2}J_{\pm\nu}(p z_1)J_{\pm\nu}(p z_2) (z_1 z_2)^{\pm\nu},
\end{align}
where $\pm\nu=\Delta_{\pm}-\frac{d}{2}$.
We define the ``Disc" operation as follows,
\begin{align}\label{eq:disc_op}
    \text{Disc}_{s}f(s^2)=f(s^2+i\epsilon)-f(s^2-i\epsilon)
\end{align}
Note that this looks a bit different than the disc operation given in \cite{Melville:2021lst}. However, these can be shown to be equivalent \cite{Das:2025qsh}. Performing the ``Disc" operation on \eqref{eq:BtB_disp_form} gives the following,
\begin{align}
    \text{Disc}_{s}G_{\pm\nu}(s,z_1,z_2)\propto(-2\pi i)J_{\pm\nu}(i s z_1)J_{\pm\nu}(i s z_2) (z_1 z_2)^{\pm\nu},
\end{align}
where we have used
\begin{align}\label{disc}
    \frac{1}{p^2+s^2+i\epsilon}-\frac{1}{p^2+s^2-i\epsilon}=-2\pi i \delta(p^2+s^2)\;.
\end{align}
Using an appropriate rescaling with the momenta, one can show that\footnote{This property is true for generic $\nu$, and we have verified this for $\nu=1/2$ which correspond to conformally coupled scalars.}, 
\begin{align}\label{eq:J_K_rel}
    J_{\pm\nu}(ix)\propto K_\nu(x)\mp K_\nu(-x)\;.
\end{align}
To show that this is equivalent to the notion discussed in section\ref{ssec:dS space unitarity}, let us look at the four-point wavefunction coefficient which for a generic scalar exchange is given as,
\begin{align}
     \psi_4(p_1,p_2,p_3,p_4,s)=(p_1p_2p_3p_4)^{1/2}\int_0^\infty dz_1dz_2 K_\nu(p_1 z_1)K_\nu(p_2 z_1)G_{\nu}(s,z_1,z_2)K_\nu(p_3 z_2)K_\nu(p_4 z_2),
\end{align}
where $s=\abs{\vec{p}_1+\vec{p}_2}$, then taking discontinuity w.r.t $s$ will give,
\begin{align}
    \text{Disc}_{s}\psi_4(p_1,p_2,p_3,p_4,s)&\propto (p_1p_2p_3p_4)^{1/2}\Big(\int_0^\infty dz_1K_\nu(p_1 z_1)K_\nu(p_2 z_1)\{K_\nu(s z_1)- K_\nu(-s z_1)\}\Big)\nonumber\\
    & \times\Big(\int_0^\infty dz_2K_\nu(p_3 z_2)K_\nu(p_4 z_2)\{K_\nu(s z_2)- K_\nu(-s z_2)\}\Big)\nonumber\\
&=\text{Disc}_{s}\psi_3(p_1,p_2,s)\text{Disc}_{s}\psi_3(p_3,p_4,s)
\end{align}
which is the same as given in equation \eqref{discwfc} upto normalisation and agrees with \cite{Melville:2021lst}. From this example, one can see that the two notions of discontinuity are equivalent.

Similarly, one can see thatbthe same notion of discontinuity produces the action of ``Disc" for the in-in correlators. For example, consider the four-point correlator due to a scalar exchange in conformally coupled $\phi^3$ theory. There are two possible exchanges allowed: (+) and (-) exchanges. The (+) exchange has the following expression
\begin{align}\label{plus}
\left\langle \phi_1(k_1) \phi_2(k_2) \phi_3(k_3) \phi_4(k_4) \right\rangle_{s}^+\propto&\;g^2\int_0^\infty dz_1dz_2 \frac{e^{-k_{12}z_1}e^{-k_{34}z_2}}{z_1z_2}\int_{-\infty}^\infty \frac{pdp}{p^2+s^2}J_{+.\frac{1}{2}}(pz_1)J_{+\frac{1}{2}}(pz_2),
\end{align}
while the (-) exchange leads to
\begin{align}\label{minus}
\left\langle \phi_1(k_1) \phi_2(k_2) \phi_3(k_3) \phi_4(k_4) \right\rangle_{s}^-\propto&\;-g^2\int_0^\infty dz_1dz_2 \frac{e^{-k_{12}z_1}e^{-k_{34}z_2}}{z_1z_2}\int_{-\infty}^\infty \frac{pdp}{p^2+s^2}J_{-\frac{1}{2}}(pz_1)J_{-\frac{1}{2}}(pz_2).
\end{align}
Using the definition of ``Disc" in \eqref{disc} and the property in \eqref{eq:J_K_rel}, the action of ``Disc" for \eqref{plus} is
\begin{align}\label{Discplus}
\textrm{Disc}_s\Big(\langle \phi_1(k_1) \phi_2(k_2) \phi_3(k_3) \phi_4(k_4) \rangle_{s}^+\Big)\propto&\;g^2\int_0^\infty dz_1dz_2 \frac{e^{-k_{12}z_1}e^{-k_{34}z_2}}{z_1z_2}(e^{-sz_1}-e^{sz_1})(e^{-sz_2}-e^{sz_2}),
\end{align}
and for \eqref{minus} we get the following expression,
\begin{align}\label{Discminus}
\textrm{Disc}_s\Big(\langle \phi_1(k_1) \phi_2(k_2) \phi_3(k_3) \phi_4(k_4) \rangle_{s}^-\Big)\propto&\;-g^2\int_0^\infty dz_1dz_2 \frac{e^{-k_{12}z_1}e^{-k_{34}z_2}}{z_1z_2}(e^{-sz_1}+e^{sz_1})(e^{-sz_2}+e^{sz_2}).
\end{align}
The in-in correlator is the sum of both $(+)$ and $(-)$ exchanges and the final expression is given by the sum of \eqref{Discplus} and \eqref{Discminus} as,
\begin{align}
    \textrm{Disc}_s\Big(\langle \phi_1(k_1) \phi_2(k_2) \phi_3(k_3) \phi_4(k_4) \rangle_{s}\Big)=&\frac{1}{2s}{\textrm{Disc}}_s\Big(\langle \phi_1(k_1) \phi_2(k_2) \phi(s)\rangle\Big){\textrm{Disc}}_s\Big(\langle \phi(s) \phi_3(k_3) \phi_4(k_4) \rangle\Big)\notag\\
-&\frac{1}{2s}\overline{\textrm{Disc}}_s\Big(\langle \phi_1(k_1) \phi_2(k_2) \phi(s)\rangle\Big)\overline{\textrm{Disc}}_s\Big(\langle \phi(s) \phi_3(k_3) \phi_4(k_4)\rangle\Big),
\end{align}
where ${\textrm{Disc}}_sf(s)=f(s)-f(-s)$ and $\overline{\textrm{Disc}}_sf(s)=f(s)+f(-s)$. This matches with the computation of ``Disc" on in-in correlators using Schwinger-Keldysh formalism \cite{Das:2025qsh,Colipi-Marchant:2025oin}.
In the following section, we will show that the effect of the ``Disc" operation on the in-in correlators is the same as taking Cutkosky cuts on the flat-space diagrams with appropriate dressing after analytic continuation \cite{Chowdhury:2025ohm}.


\section{Lifting flat-space unitarity to dS}\label{sec:lifting}
In this section,  we will show that the discontinuity relations in EAdS are the result of optical theorem in flat-space when dressed appropriately using the dressing rules given in section \ref{sec:dressing}. In particular we will take the relation
\begin{align}\label{eq:TTdagger_rel1}
    -i(\mathcal{T}-\mathcal{T}^\dagger)=\mathcal{T}\mathcal{T}^\dagger\;.
\end{align} and uplift it (after analytic continuation) to dS space using dressing rules and show that this gives precisely the cutting rules for in-in correlators as given below and also discussed in section.\,\ref{ssec:dS space unitarity}.
\begin{align}\label{disccorr1}
\mathrm{Disc}_p \mathcal{B}^{(2)}(\{k_L,k_R\};p)
=
\frac{1}{2 P_p(\eta_0)}
\Big(&
\mathrm{Disc}_p \mathcal{B}^{(1)}(\{k_L\},p)
\times
\mathrm{Disc}_p {\mathcal{B}}^{(1)}(\{k_R\},p)\notag\\
&-
\widetilde{
\mathrm{Disc}}_p \tilde{\mathcal{B}}^{(1)}(\{k_L\},p)
\times
\widetilde{
\mathrm{Disc}}_p \tilde{\mathcal{B}}^{(1)}(\{k_R\},p)
\Big),
\end{align}

We show this for tree and loop level processes involving conformally coupled scalars with polynomial interactions, but it can be easily generalizable for generic interaction involving other fields.
We start with the simple case involving scalar correlators. We will first look at tree level correlators and then move onto loops.

\subsection{Tree level scalar correlators}
We start with the s-channel exchange of the four point amplitude for $\phi^3$ theory which is given as,
\begin{align}
\mathcal{A}_4^{\textrm{Tree}}&=g^2\int_{-\infty}^\infty d^4k\frac{1}{k^2}\delta^4(k_1+k_2-k)\delta^4(k_3+k_4+k)=\frac{g^2}{(p^2-s^2)}.
\end{align}
where $p=|\vec{k}_1|+|\vec{k}_2|$ is the energy and $\vec{s}=\vec{k}_1+\vec{k}_2$ is the spatial momentum. Taking the imaginary part of the flat-space propagator amounts to putting the internal leg on-shell \cite{Peskin:1995ev,Schwartz:2014sze},
\begin{align}\label{disctree}
2\textrm{Im}\mathcal{A}_4^{\textrm{Tree}}=g^2\delta(p^2-s^2).  
\end{align}
The RHS of optical theorem \eqref{eq:matrix for TT 2 part} is given by
\begin{align}\label{rhstree}
\int\frac{d^3q}{2|\vec{q}|}|\mathcal{A}_3(s)|^2\delta^4(k_1+k_2-q)=\frac{g^2}{2p}\delta(p-|\vec{s}|),
\end{align}
where the three point amplitude $\mathcal{A}_3(s)=g$. Therefore, we see that, 
\begin{align}\label{eq:opt_theorem_tree}
2\textrm{Im}\mathcal{A}_4^{\textrm{Tree}}=\int\frac{d^3q}{2|\vec{q}|}|\mathcal{A}_3(s)|^2\delta^4(k_1+k_2-q)\;,
\end{align}
and hence the optical theorem as given in equation \eqref{eq:matrix for TT 2 part} is satisfied. When lifting the flat space amplitude to EAdS, we analytically continue the $s$ variable as $s\rightarrow is$. We will use this convention whenever we lift the flat-space amplitude to get EAdS correlators \cite{Chowdhury:2025ohm}.

After analytic continuation, we now dress both sides of \eqref{eq:opt_theorem_tree} with the appropriate dressing factors to get the analog of optical theorem for in-in correlators in EAdS.  As shown in section \ref{sec:dressing}, we have to dress the flat-space amplitude with both the $(+)$ and $($-$)$ dressing factors as given in \eqref{phi3tree+} and \eqref{phi3tree-}. Applying the dressing for $(+)$ exchange on the LHS of \eqref{eq:opt_theorem_tree}, we get
\begin{align}\label{discintold}
&g^2\int_{-\infty}^\infty dp\bigg(\int_0^\infty dz_1dz_2 \frac{e^{-k_{12}z_1}e^{-k_{34}z_2}}{z_1z_2} \textrm{sin}(pz_1)\textrm{sin}(pz_2)\bigg)\delta(p^2+s^2)\nonumber\\
=&g^2\int_0^\infty dz_1dz_2 K_{k_1}(z_1)K_{k_2}(z_1)\text{Disc}_{s}G(s,z_1,z_2)K_{k_3}(z_2)K_{k_4}(z_2),
\end{align}
where we have interpreted the delta function as the ``Disc" operation defined through \eqref{eq:disc_op} and \eqref{disc}, and we have chosen to close the contour in the upper half plane. Thus, the action of ``Disc" on the full correlator is defined as
\begin{align}\label{discint}
\textrm{Disc}_s\Big(\langle \phi_1(k_1) \phi_2(k_2) \phi_3(k_3) \phi_4(k_4) \rangle_{s}^+\Big)=g^2\int_{-\infty}^\infty dp\bigg(\int_0^\infty dz_1dz_2 \frac{e^{-k_{12}z_1}e^{-k_{34}z_2}}{z_1z_2} \textrm{sin}(pz_1)\textrm{sin}(pz_2)\bigg)\delta(p^2+s^2).
\end{align}
Performing the same dressing on the RHS of \eqref{rhstree}, we obtain
\begin{align}\label{eq:DiscDisc treeold}
&g^2\int_{-\infty}^\infty \frac{dp}{2p}\Bigg(\int_0^\infty dz_1 \frac{e^{-k_{12}z_1}\textrm{sin}(pz_1)}{z_1} \Bigg)\Bigg(\int_0^\infty dz_2 \frac{e^{-k_{34}z_2}\textrm{sin}(pz_2)}{z_2} \Bigg)\delta(p-is)\nonumber\\
=&\frac{1}{2s}\Big(g\int_0^\infty dz_1\frac{e^{-k_{12}z_1}(e^{sz_1}-e^{-sz_1})}{z_1}\Big)\Big(g\int_0^\infty dz_2\frac{e^{-k_{34}z_2} (e^{sz_2}-e^{-sz_2})}{z_2}\Big)\notag\\
=&\frac{1}{2s}\textrm{Disc}_s\Big(\langle \phi_1(k_1) \phi_2(k_2) \phi(s)\rangle\Big)\textrm{Disc}_s\Big(\langle \phi(s) \phi_3(k_3) \phi_4(k_4)\rangle\Big).
\end{align}
where ${\textrm{Disc}}_sf(s)=f(s)-f(-s).$ Therefore, equations \eqref{eq:opt_theorem_tree}, \eqref{discint} and \eqref{eq:DiscDisc treeold} imply that
\begin{align}\label{eq:DD plus}
\textrm{Disc}_s\Big(\langle \phi_1(k_1) \phi_2(k_2) \phi_3(k_3) \phi_4(k_4) \rangle_{s}^+\Big)
=&\frac{1}{2s}\textrm{Disc}_s\Big(\langle \phi_1(k_1) \phi_2(k_2) \phi(s)\rangle\Big)\textrm{Disc}_s\Big(\langle \phi(s) \phi_3(k_3) \phi_4(k_4)\rangle\Big).
\end{align}
Similarly, for the $($-$)$ exchange we get,
\begin{align}\label{eq:four_mm_discb_factor}
\textrm{Disc}_s\Big(\langle \phi_1(k_1) \phi_2(k_2) \phi_3(k_3) \phi_4(k_4) \rangle_{s}^-\Big)
=-&\frac{1}{2s}\overline{\textrm{Disc}}_s\Big(\langle \phi_1(k_1) \phi_2(k_2) \phi(s)\rangle\Big)\overline{\textrm{Disc}}_s\Big(\langle \phi(s) \phi_3(k_3) \phi_4(k_4) \rangle\Big),
\end{align}
where we define $\overline{\textrm{Disc}}_sf(s)=f(s)+f(-s)$. Adding \eqref{eq:DD plus} and \eqref{eq:four_mm_discb_factor} gives the discontinuity of the full in-in correlator as,
\begin{align}\label{discphi3tree}
\textrm{Disc}_s\Big(\langle \phi_1(k_1) \phi_2(k_2) \phi_3(k_3) \phi_4(k_4) \rangle_{s}\Big)=&\frac{1}{2s}{\textrm{Disc}}_s\Big(\langle \phi_1(k_1) \phi_2(k_2) \phi(s)\rangle\Big){\textrm{Disc}}_s\Big(\langle \phi(s) \phi_3(k_3) \phi_4(k_4) \rangle\Big)\notag\\
-&\frac{1}{2s}\overline{\textrm{Disc}}_s\Big(\langle \phi_1(k_1) \phi_2(k_2) \phi(s)\rangle\Big)\overline{\textrm{Disc}}_s\Big(\langle \phi(s) \phi_3(k_3) \phi_4(k_4)\rangle\Big).
\end{align}
This result is independently derived in dS space \cite{Das:2025qsh}, by taking discontinuities on the in-in correlator. Hence, we see that optical theorem in flat-space implies the discontinuity relations in EAdS.

\subsubsection*{General cutting rule for two-site exchange diagrams}
Let us see how equation \eqref{discphi3tree} can help us in generalising to any n-point function with 2 sites(2 vertices) connected with one propagator. Note that equation \eqref{discphi3tree} allowed for both the possible exchanges i.e. (+) and ($-$) at tree level diagram for two vertices and therefore the results obtained can be straightforwardly generalised to higher point. Consider a generic two site correlator at tree level for conformally coupled scalar with polynomial interaction
\begin{align}
\langle\Phi(\{k_L\})\Phi(\{k_R\})\rangle,
\end{align}
where $\Phi(\{k_L\})$ and $\Phi(\{k_R\})$ denote the set of external fields connecting to the left and right vertex respectively and $s$ is the energy associated with the left vertex. The action of ``Disc" operation the correlator will thus yield
\begin{align}\label{eq:disc most general}
\textrm{Disc}_s\Big(\langle\Phi(\{k_L\})\Phi(\{k_R\}),s\rangle\Big)&=\frac{1}{2s}{\textrm{Disc}}_s\Big(\langle \Phi(\{k_L\}) \phi(s)\rangle\Big){\textrm{Disc}}_s\Big(\langle \phi(s) \Phi(\{k_R\}) \rangle\Big)\notag\\
&-\frac{1}{2s}\overline{\textrm{Disc}}_s\Big(\langle \Phi(\{k_L\}) \phi(s)\rangle\Big)\overline{\textrm{Disc}}_s\Big(\langle \phi(s)\Phi(\{k_R\})\rangle\Big).
\end{align}
This matches with result presented in \cite{Das:2025qsh}. We now move on to the cases involving loops.

\subsection{Loops for scalar correlators}
Let us now discuss the effects of uplifting flat space discontinuity of loop diagrams to their de Sitter counterparts. We begin with the $\phi^4$ theory.

\subsection*{$\mathbf{\phi}^4$ theory}
Consider the case of 1-loop four-point amplitude with $\phi^4$ interaction. The optical theorem as given in equation \eqref{eq:matrix for TT 2 part} states the following,
\begin{align}\label{phi4looprhs}
   2 \text{Im}( \mathcal{A}_4^{1-\textrm{loop}}(s))&=\frac{\lambda^2}{2}\int \frac{d^3l_1}{2|\vec{l}_1|}\frac{d^3l_2}{2|\vec{l}_2|} |\mathcal{A}_4(s)|^2\delta(k_1+k_2+p_1-p_2)\delta^3(\vec{k}_1+\vec{k}_2+\vec{l}_1-\vec{l}_2)
\end{align}
where $|\mathcal{A}_4(s)|^2=\lambda^2$ and $|\vec{l}_i|$ are phase space integrals over exchange particles. The one loop four point function is given as,
\begin{align}
    \mathcal{A}_4^{1-\textrm{loop}}(s)=\frac{\lambda^2}{2}\int d^4L_1 d^4L_2\frac{1}{L_1^2 L_2^2}\delta^4(K_1+K_2+L_1-L_2),
\end{align}
where we have defined the following Euclidean 4-vectors $K_i=(k_i,\vec{k}_i)$.
Taking the imaginary part of the amplitude is the same as putting both the internal propagators on-shell. This gives the following,
\begin{align}\label{phi4looplhs}
    2 \text{Im}( \mathcal{A}_4^{1-\textrm{loop}}(s))=&\frac{\lambda^2}{2}\int d^4L_1 d^4L_2\;\delta(L_1^2)\delta(L_2^2)\delta^4(K_1+K_2+L_1-L_2)\notag\\
    =&\frac{\lambda^2}{2}\int d^3l_1\int d^3l_2\int dp_1\int dp_2\ \delta(p_1^2+|\vec{l}_1|^2)\delta(p_2^2+|\vec{l}_2|^2)\delta^4(K_1+K_2+L_1-L_2).
\end{align}
which agrees with equation \eqref{phi4looprhs}.

To get the corresponding dS correlators, we relax the energy conservation and dress the amplitudes with appropriate dressing factor for (++) exchange and ($--$) exchange as given in \eqref{phi4loop+} and \eqref{phi4loop-} respectively. Moreover, from now on, we will work only with the integrands without the $d^3l$ integration, which will be defined as follows
\begin{align}
\left\langle \phi_1(k_1) \phi_2(k_2) \phi_3(k_3) \phi_4(k_4) \right\rangle=\int d^3l_1\int d^3l_2 \;\mathcal{I}\left\langle \phi_1(k_1) \phi_2(k_2) \phi_3(k_3) \phi_4(k_4) \right\rangle.
\end{align}
Starting from the LHS of \eqref{phi4looprhs}, which is explicitly given in \eqref{phi4looplhs}, the dressed version of the integrand for the (++) exchange gives,
\begin{align}\label{LHS++}
&\lambda^2\int dp_1dp_2\int_0^\infty dz_1dz_2 e^{-k_{12}z_1} \textrm{sin}(p_1z_1) \textrm{sin}(p_2z_1)e^{-k_{34}z_2}\textrm{sin}(p_1z_2)\textrm{sin}(p_2z_2)\delta(p_1^2+|\vec{l}_1|^2)\delta(p_2^2+|\vec{l}_2|^2)\notag\\
&=\textrm{Disc}_{l_1}\textrm{Disc}_{l_2}\Big(\mathcal{I}\left\langle \phi_1(k_1) \phi_2(k_2) \phi_3(k_3) \phi_4(k_4) \right\rangle_{s}^{++}\Big).
\end{align}
Similarly, the dressed version of the integrand for ($--$) exchange leads to 
\begin{align}\label{LHS--}
&\lambda^2\int dp_1dp_2 \int_0^\infty dz_1 e^{-k_{12}z_1} \textrm{cos}(p_1z_1) \textrm{cos}(p_2z_1)
\int_0^\infty dz_2e^{-k_{34}z_2}\textrm{cos}(p_1z_2)\textrm{cos}(p_2z_2)\delta(p_1^2+|\vec{l}_1|^2)\delta(p_2^2+|\vec{l}_2|^2)\notag\\
&=\textrm{Disc}_{l_1}\textrm{Disc}_{l_2}\Big(\mathcal{I}\left\langle \phi_1(k_1) \phi_2(k_2) \phi_3(k_3) \phi_4(k_4) \right\rangle_{s}^{--}\Big).
\end{align}
Doing the same procedure for (++) exchange for the RHS \eqref{phi4looprhs} yields
\begin{align}\label{RHS++}
&\frac{\lambda^2}{2}\int\frac{dp_1dp_2}{4p_1p_2}\int_0^\infty dz_1 e^{-k_{12}z_1} \textrm{sin}(p_1z_1) \textrm{sin}(p_2z_1)\int_0^\infty dz_2e^{-k_{34}z_2}\textrm{sin}(p_1z_2)\textrm{sin}(p_2z_2)\delta(p_1-i|\vec{l}_1|)\delta(p_2-i|\vec{l}_2|)\notag\\
&=\frac{1}{4|\vec{l}_1||\vec{l}_2|}\Big(\textrm{Disc}_{l_1}\textrm{Disc}_{l_2}\langle\phi(k_1)\phi(k_2)\phi(|\vec{l}_1|)\phi(|\vec{l}_2|)\rangle\Big)\Big(\textrm{Disc}_{l_1}\textrm{Disc}_{l_2}\langle\phi(|\vec{l}_1|)\phi(|\vec{l}_2|)\phi(k_3)(\phi(k_4)\rangle\Big),
\end{align}
while for ($--$) exchange it results in
\begin{align}\label{RHS--}
&\frac{\lambda^2}{2}\int\frac{dp_1dp_2}{4p_1p_2}\int_0^\infty dz_1 e^{-k_{12}z_1} \textrm{cos}(p_1z_1) \textrm{cos}(p_2z_1)\int_0^\infty dz_2e^{-k_{34}z_2}\textrm{cos}(p_1z_2)\textrm{cos}(p_2z_2)\delta(p_1-i|\vec{l}_1|)\delta(p_2-i|\vec{l}_2|)\notag\\
&=\frac{1}{4|\vec{l}_1||\vec{l}_2|}\Big(\overline{\textrm{Disc}}_{l_1}\overline{\textrm{Disc}}_{l_2}\langle\phi(k_1)\phi(k_2)\phi(|\vec{l}_1|)\phi(|\vec{l}_2|)\rangle\Big)\Big(\overline{\textrm{Disc}}_{l_1}\overline{\textrm{Disc}}_{l_2}\langle\phi(|\vec{l}_1|)\phi(|\vec{l}_2|)\phi(k_3)(\phi(k_4)\rangle\Big).
\end{align}
Using the dressed version of \eqref{phi4looprhs}, along with \eqref{LHS++}, \eqref{LHS--}, \eqref{RHS++} and \eqref{RHS--} leads to the following equality
\begin{align}\label{phi4loopdisc}
&\textrm{Disc}_{l_1}\textrm{Disc}_{l_2}\Big(\mathcal{I}\left\langle \phi_1(k_1) \phi_2(k_2) \phi_3(k_3) \phi_4(k_4) \right\rangle_{s}\Big)\notag\\
&=\textrm{Disc}_{l_1}\textrm{Disc}_{l_2}\Big(\mathcal{I}\left\langle \phi_1(k_1) \phi_2(k_2) \phi_3(k_3) \phi_4(k_4) \right\rangle_{s}^{++}\Big)+\textrm{Disc}_{l_1}\textrm{Disc}_{l_2}\Big(\mathcal{I}\left\langle \phi_1(k_1) \phi_2(k_2) \phi_3(k_3) \phi_4(k_4) \right\rangle_{s}^{--}\Big)\notag\\
&=\frac{1}{4|\vec{l}_1||\vec{l}_2|}\Big(\textrm{Disc}_{l_1}\textrm{Disc}_{l_2}\langle\phi(k_1)\phi(k_2)\phi(|\vec{l}_1|)\phi(|\vec{l}_2|)\rangle\Big)\Big(\textrm{Disc}_{l_1}\textrm{Disc}_{l_2}\langle\phi(|\vec{l}_1|)\phi(|\vec{l}_2|)\phi(k_3)(\phi(k_4)\rangle\Big)\notag\\
&+\frac{1}{4|\vec{l}_1||\vec{l}_2|}\Big(\overline{\textrm{Disc}}_{l_1}\overline{\textrm{Disc}}_{l_2}\langle\phi(k_1)\phi(k_2)\phi(|\vec{l}_1|)\phi(|\vec{l}_2|)\rangle\Big)\Big(\overline{\textrm{Disc}}_{l_1}\overline{\textrm{Disc}}_{l_2}\langle\phi(|\vec{l}_1|)\phi(|\vec{l}_2|)\phi(k_3)(\phi(k_4)\rangle\Big).
\end{align}
which is in agreement with \cite{Das:2025qsh}, which was independently computed using in-in formalism.

\subsection*{$\mathbf{\phi}^3$ theory}
Consider now the case of 1-loop correction to the propagator for $\phi^3$ interaction, where \eqref{eq:TTdagger_rel} takes the following form
\begin{align}\label{eq:A1loop_Amid_rel}
   2 \text{Im}( \mathcal{A}_2^{1-\textrm{loop}}(s))=\frac{1}{2}\int \frac{d^3l_1}{2|\vec{l}_1|}\frac{d^3l_2}{2|\vec{l}_2|}|\mathcal{A}(s)|^2\delta(k_1+p_1-p_2)\delta^3(\vec{k}_1+\vec{l}_1-\vec{l}_2)
\end{align}
where $|\mathcal{A}(s)|^2=g^2$. The LHS of \eqref{eq:A1loop_Amid_rel} is given as follows
\begin{align}\label{eq:ImA}
    2 \text{Im}( \mathcal{A}_2^{1-\textrm{loop}}(s))=&g^2\int d^4L_1 d^4L_2\;\delta(L_1^2)\delta(L_2^2)\delta^4(K_1+L_1-L_2).
\end{align}
Again to get the EAdS correlators, we perform the same steps as before and dress \eqref{eq:A1loop_Amid_rel} and \eqref{eq:ImA} with the dressing factors given in \eqref{phi3loop}. Performing the same drill as in the earlier cases, we obtain the following expression after adding for all possible exchanges
\begin{align}\label{2site1loop}
&\textrm{Disc}_{l_1}\textrm{Disc}_{l_2}\Big(\mathcal{I}\left\langle \phi_1(k_1) \phi_2(k_2) \right\rangle\Big)\notag\\
&=\frac{1}{4|\vec{l}_1||\vec{l}_2|}\Big(\textrm{Disc}_{l_1}\textrm{Disc}_{l_2}\langle\phi(k_1)\phi(|\vec{l}_1|)\phi(|\vec{l}_2|)\rangle\Big)\Big(\textrm{Disc}_{l_1}\textrm{Disc}_{l_2}\langle\phi(|\vec{l}_1|)\phi(|\vec{l}_2|)\phi(k_2)\rangle\Big)\notag\\
&+\frac{1}{4|\vec{l}_1||\vec{l}_2|}\Big(\overline{\textrm{Disc}}_{l_1}\overline{\textrm{Disc}}_{l_2}\langle\phi(k_1)\phi(|\vec{l}_1|)\phi(|\vec{l}_2|)\rangle\Big)\Big(\overline{\textrm{Disc}}_{l_1}\overline{\textrm{Disc}}_{l_2}\langle\phi(|\vec{l}_1|)\phi(|\vec{l}_2|)\phi(k_2)\rangle\Big)\notag\\
&-\frac{1}{4|\vec{l}_1||\vec{l}_2|}\Big(\overline{\textrm{Disc}}_{l_1}\textrm{Disc}_{l_2}\langle\phi(k_1)\phi(|\vec{l}_1|)\phi(|\vec{l}_2|)\rangle\Big)\Big(\overline{\textrm{Disc}}_{l_1}\textrm{Disc}_{l_2}\langle\phi(|\vec{l}_1|)\phi(|\vec{l}_2|)\phi(k_2)\rangle\Big)\notag\\
&-\frac{1}{4|\vec{l}_1||\vec{l}_2|}\Big(\textrm{Disc}_{l_1}\overline{\textrm{Disc}}_{l_2}\langle\phi(k_1)\phi(|\vec{l}_1|)\phi(|\vec{l}_2|)\rangle\Big)\Big(\textrm{Disc}_{l_1}\overline{\textrm{Disc}}_{l_2}\langle\phi(|\vec{l}_1|)\phi(|\vec{l}_2|)\phi(k_2)\rangle\Big).
\end{align}
This is again in agreement with  \cite{Das:2025qsh}. Thus we observe that the de Sitter cutting rule can be obtained by dressing the Cutkosky rules by appropriate dressing factors.

\subsection*{General cutting rule for two-site one-loop diagrams}
As we saw for the case of tree level correlator, we can generalise the cuts for any 2 site (2 vertices) 1-loop diagrams. Since $\phi^3$ theory allowed for all the possible exchanges for the 1-loop diagram for two vertices. Inspired by this result, consider a generic two site correlator at one loop for conformally coupled scalar with polynomial interaction
\begin{align}
\langle\Phi(\{k_L\})\Phi(\{k_R\})\rangle,
\end{align}
where $\Phi(\{k_L\})$ and $\Phi(\{k_R\})$ denote the set of external fields connecting to the left and right vertex respectively. The action of ``Disc" operation on the integrand will yield
\begin{align}\label{eq:disc most general}
&\textrm{Disc}_{l_1}\textrm{Disc}_{l_2}\Big(\mathcal{I}\langle\Phi(\{k_L\})\Phi(\{k_R\}),l_1,l_2\rangle\Big)\notag\\
&=\frac{1}{4|\vec{l}_1||\vec{l}_2|}\Big(\textrm{Disc}_{l_1}\textrm{Disc}_{l_2}\langle\Phi(\{k_L\})\phi(|\vec{l}_1|)\phi(|\vec{l}_2|)\rangle\Big)\Big(\textrm{Disc}_{l_1}\textrm{Disc}_{l_2}\langle\phi(|\vec{l}_1|)\phi(|\vec{l}_2|)\Phi(\{k_R\})\rangle\Big)\notag\\
&+\frac{1}{4|\vec{l}_1||\vec{l}_2|}\Big(\overline{\textrm{Disc}}_{l_1}\overline{\textrm{Disc}}_{l_2}\langle\Phi(\{k_L\})\phi(|\vec{l}_1|)\phi(|\vec{l}_2|)\rangle\Big)\Big(\overline{\textrm{Disc}}_{l_1}\overline{\textrm{Disc}}_{l_2}\langle\phi(|\vec{l}_1|)\phi(|\vec{l}_2|)\Phi(\{k_R\})\rangle\Big)\notag\\
&-\frac{1}{4|\vec{l}_1||\vec{l}_2|}\Big(\overline{\textrm{Disc}}_{l_1}\textrm{Disc}_{l_2}\langle\Phi(\{k_L\})\phi(|\vec{l}_1|)\phi(|\vec{l}_2|)\rangle\Big)\Big(\overline{\textrm{Disc}}_{l_1}\textrm{Disc}_{l_2}\langle\phi(|\vec{l}_1|)\phi(|\vec{l}_2|)\Phi(\{k_R\})\rangle\Big)\notag\\
&-\frac{1}{4|\vec{l}_1||\vec{l}_2|}\Big(\textrm{Disc}_{l_1}\overline{\textrm{Disc}}_{l_2}\langle\Phi(\{k_L\})\phi(|\vec{l}_1|)\phi(|\vec{l}_2|)\rangle\Big)\Big(\textrm{Disc}_{l_1}\overline{\textrm{Disc}}_{l_2}\langle\phi(|\vec{l}_1|)\phi(|\vec{l}_2|)\Phi(\{k_R\})\rangle\Big).
\end{align}
Thus we claim that \eqref{eq:disc most general} is the most general discontinuity allowed for 1-loop diagram for two vertices for any polynomial interaction, with \eqref{phi4loopdisc} and \eqref{2site1loop} being a special case.

\section{Conclusion}\label{sec:conc}
In this paper, we have presented a direct route to cosmological cutting rules: start from the flat-space optical theorem/Cutkosky rules, analytically continue to EAdS/dS kinematics, and apply the cosmological dressing map. In this representation, the flat-space on-shell delta functions uplift into the discontinuity operations acting on the exchanged energy variable in EAdS/dS, yielding a practical diagram-by-diagram dictionary for cosmological cutting rules.

This  viewpoint also explains why correlator-level discontinuity relations can involve auxiliary combinations rather than correlators alone: these are simply what the dressed “cut = discontinuity” statement produces once the correct in-in/EAdS analytic structure is imposed. We expect this framework to help organize loop computations. {Moreover, this may also lead to a non-perturbative framework for investigating dS unitarity}. 

More broadly, this dictionary suggests a natural pathway for uplifting flat-space amplitude results to de Sitter space, including BCFW recursion/constructibility ideas, generalized unitarity for wavefunction coefficients and correlators, dispersion relations and sum rules motivated by analyticity and causality, positivity-type constraints in suitable regimes, and the systematic use of soft limits and cosmological consistency relations as bootstrap inputs \cite{Baumann:2022jpr}. It would also be interesting to explore how familiar double-copy \cite{Adamo:2022dcm,Armstrong:2020woi,Ansari:2025fvi}–like patterns for example in four point functions—are reflected after dressing.

Several extensions of this work are immediate, including a generalization to spinning fields. Since the propagators for spinning fields coincide with those of conformally coupled scalars up to momentum rescaling, such extensions are expected to be relatively straightforward. Other extensions include generic massive scalars, departures from exact de Sitter such as inflationary correlators, and more general loop topologies.

\acknowledgments
The authors would like to thank N. Kundu for useful discussions. SJ would like to thank Lodha Mathematical Sciences Insititute (LMSI) and Tata Institute of Fundamental Research (TIFR) for organising the workshop ``Discussion on Quantum Spacetime" where parts of the work was completed. AA and DM would like to thank the organisers of the ``20th Asian Winter School (AWS)" for the hospitality where parts of the project was finished. AA acknowledges the support from a Senior Research Fellowship,
granted by the Human Resource Development Group, Council of Scientific and Industrial
Research, Government of India. We would especially like to acknowledge our debt to the people of India for their steady support of research in basic sciences.

\bibliographystyle{JHEP}
\bibliography{biblio}

\end{document}